\documentclass[12pt]{article}

\usepackage{amsfonts, amsmath}

\newcommand{\be}{\begin{equation}} \newcommand{\ee}{\end{equation}}

\begin{document}
\title{Non-Unitary and Unitary Transitions
 in Generalized Quantum  Mechanics, New Small Parameter
 and Information Problem Solving} \thispagestyle{empty}

\author{A.E.Shalyt-Margolin\hspace{1.5mm}\thanks
{Fax: (+375) 172 326075; e-mail: a.shalyt@mail.ru;alexm@hep.by}}
\date{}
\maketitle
 \vspace{-25pt}
{\footnotesize\noindent  National Center of Particles and High
Energy Physics, Bogdanovich Str. 153, Minsk 220040, Belarus\\
{\ttfamily{\footnotesize
\\ PACS: 03.65; 05.20
\\
\noindent Keywords: density matrix deformation,Heisenberg algebra
deformation, unitarity, information
            paradox problem}}

\rm\normalsize \vspace{0.5cm}
\begin{abstract}
Quantum Mechanics of the Early Universe is considered as
deformation of a well-known Quantum Mechanics. Similar to previous
works of the author, the principal approach is based on
deformation of the density matrix with concurrent development of
the wave function deformation in the respective Schr{\"o}dinger
picture, the associated deformation parameter being interpreted as
a new small parameter. It is demonstrated that the existence of
black holes in the suggested approach in the end twice causes
nonunitary transitions resulting in the unitarity. In parallel
this problem is considered in other terms: entropy density,
Heisenberg algebra deformation terms, respective deformations of
Statistical Mechanics, - all showing the identity of the basic
results. From this an explicit solution for Hawking's informaion
paradox has been derived.
\end{abstract}
\newpage
\section{Introduction}
As is known, the Early Universe Quantum Mechanics (Quantum
Mechanics at Planck scale) is distinguished from a well- known
Quantum Mechanics at conventional scales \cite{r1},\cite{r2}by the
fact that in the first one the Generalized Uncertainty Relations
(GUR) are fulfilled, resulting in the emergence of a fundamental
length, whereas in the second one the usual Heisenberg Uncertainty
Relations are the case. In case of Quantum Mechanics with
Fundamental Length (QMFL) all three well-known fundamental
constants are involved $G$,$c$ è $\hbar$, while the classical QM
is associated only with a single one $\hbar$. It is obvious that
transition from the first to the second one within the inflation
expansion is a nonunitary process, i.e. the process where the
probabilities are not retained \cite{r3}, \cite{r4}. Because of
this, QMFL is considered as a deformation of QM. The deformation
in Quantum Mechanics at Planck scale takes different paths:
commutator deformation or more precisely deformation of the
respective Heisenberg algebra \cite{r5},\cite{r6},\cite{r7} , i.e.
the density matrix deformation approach,developed by the author
with co-workers in a number of papers
\cite{r3},\cite{r4},\cite{r8},\cite{r9},\cite{r10}. The first
approach suffers from two serious disadvantages: 1) the
deformation parameter is a dimensional variable $\kappa$ with a
dimension of mass \cite{r5}; 2) in the limiting transition to QM
this parameter goes to infinity and fluctuations of other values
are hardly sensitive to it. Being devoid of the above limitation,
the second approach by the author's opinion is intrinsic for QMFL:
with it in QMFL the deformation parameter is represented by the
dimensionless quantity $\alpha=l_{min}^{2}/x^{2}$, where $x$ is
the measuring scale and the variation interval $\alpha$ is finite
$0<\alpha\leq1/4$ \cite{r3}, \cite{r4},\cite{r10}. Besides, this
approach contributes to the solution of particular problems such
as the information paradox problem of black holes \cite{r3} and
also the problem of an extra term in Liouville equation
\cite{r8},\cite{r9},\cite{r10}, derivation of Bekenstein-Hawking
formula from the first principles \cite{r10}, hypothesis of cosmic
censorship \cite{r9},\cite{r10}, more exact definition and
expansion of the entropy notion through the introduction of the
entropy density per minimum unit area
\cite{r9},\cite{r10},\cite{r11}. Moreover, it is demonstrated that
there exists a complete analogy in the construction and properties
of quantum mechanics and statistical density matrices at Planck
scale (density pro-matrices). It should be noted that an ordinary
statistical density matrix appears in the low-temperature limit
(at temperatures much lower than the Planck's)\cite{r12}. In the
present work the unitarity problem for QMFL is considered on the
basis of this approach. It is shown that as distinct from
Hawking's approach, in this treatment the existence of black holes
is not the  reason for the unitarity violation, rather being
responsible for its recovery. First after the Big Bang (Original
Singularity)expansion of the Universe is associated with the
occurrence of a nonunitary transition from QMFL to QM, and with
trapping of the matter by the black hole (Black Hole Singularity)
we have a reverse nonunitary process from QM to QMFL. In such a
manner a complete transition process from QMFL to the unitarity
may be recovered. This fact may be verified differently by the
introduction of a new value - entropy density matrix \cite{r11}
with demonstration of the identity of entropy densities in the
vicinity of the initial and final singularities. Thus, the
existence of black holes contributes to the reconstruction of a
symmetry of the general picture. Similar results may be obtained
in terms of the Heisenberg's algebra deformation \cite{r5} and
corresponding deformation of Statistical Mechanics
\cite{r12},\cite{r20} associated with the Generalized Uncertainty
Relations in Thermodynamics \cite{r20},\cite{r21},\cite{r22}. So
the problem of Hawking information paradox is solved by the
proposed approach: the information quantity in the Universe is
preserved. In final sections of the paper it is demonstrated that
$\alpha$ interpreted as a new small parameter has particular
advantages. This paper is a summing-up of the tentative results
obtained by the author on the information paradox as an extension
of the earlier works \cite{r3},\cite{r11} and \cite{r25}.

\section {Some Preliminary Facts}

In this section the principal features of QMFL construction are
briefly outlined first in terms of the density matrix
deformation(von Neumann's picture) and subsequently in terms of
the wave function deformation (Schr{\"o}dinger picture) \cite{r3},
\cite{r4},\cite{r9},\cite{r10}. As mentioned above, for the
fundamental deformation parameter we use $\alpha = l_{min}^{2
}/x^{2}$, where $x$ is the measuring scale.
\\
\\
\noindent {\bf Definition 1.} {\bf(Quantum Mechanics with
Fundamental Length [for Neumann's picture])}
\\
\\
\noindent Any system in QMFL is described by a density pro-matrix
of the form $${\bf
\rho(\alpha)=\sum_{i}\omega_{i}(\alpha)|i><i|},$$ where
\begin{enumerate}
\item $0<\alpha\leq1/4$;
\item Vectors $|i>$ form a full orthonormal system;
\item $\omega_{i}(\alpha)\geq 0$ and for all $i$  the
finite limit $\lim\limits_{\alpha\rightarrow
0}\omega_{i}(\alpha)=\omega_{i}$ exists;
\item
$Sp[\rho(\alpha)]=\sum_{i}\omega_{i}(\alpha)<1$,
$\sum_{i}\omega_{i}=1.$;
\item For every operator $B$ and any $\alpha$ there is a
mean operator $B$ depending on  $\alpha$:\\
$$<B>_{\alpha}=\sum_{i}\omega_{i}(\alpha)<i|B|i>.$$

\end{enumerate}should be fulfilled:
\begin{equation}\label{U1}
Sp[\rho(\alpha)]-Sp^{2}[\rho(\alpha)]\approx\alpha.
\end{equation}
Consequently we can find the value for $Sp[\rho(\alpha)]$ satisfying the
condition of Definition 1:
\begin{equation}\label{U2}
Sp[\rho(\alpha)]\approx\frac{1}{2}+\sqrt{\frac{1}{4}-\alpha}.
\end{equation}

According to point 5), $<1>_{\alpha}=Sp[\rho(\alpha)]$. Therefore
for any scalar quantity $f$ we have $<f>_{\alpha}=f
Sp[\rho(\alpha)]$. We denote the limit
$\lim\limits_{\alpha\rightarrow 0}\rho(\alpha)=\rho$ as the
density matrix. Evidently, in the limit $\alpha\rightarrow 0$ we
return to QM.

\renewcommand{\theenumi}{\Roman{enumi}}
\renewcommand{\labelenumi}{\theenumi.}

\renewcommand{\labelenumii}{\theenumii.}

In \cite{r3},\cite{r9},\cite{r10} it was shown that,

\begin{enumerate}
\item the above limit covers both Quantum
and Classical Mechanics.
\item Density pro-matrix $\rho(\alpha)$ tests singularities.
As a matter of fact, the deformation parameter $\alpha$
should assume value $0<\alpha\leq1$.  However, as seen from
(\ref{U2}), $Sp[\rho(\alpha)]$ is well defined only for
$0<\alpha\leq1/4$, i.e. for $x=il_{min}$ and $i\geq 2$ we have no
problems at all. At the point, where $x=l_{min}$ (that corresponds
to a singularity of space), $Sp[\rho(\alpha)]$ takes the complex
values.
\item It is possible to read equation (\ref{U1}) as
\begin{equation}\label{U3}
Sp[\rho(\alpha)]-Sp^{2}[\rho(\alpha)]=\alpha+a_{0}\alpha^{2}
+a_{1}\alpha^{3}+...
\end{equation}
Then for example, one of the solutions of (\ref{U1}) is
\begin{equation}\label{U4}
\rho^{*}(\alpha)=\sum_{i}\alpha_{i} exp(-\alpha)|i><i|,
\end{equation}
where all
$\alpha_{i}>0$ are independent of $\alpha$  and their sum is equal
to 1 . In this way $Sp[\rho^{*}(\alpha)]=exp(-\alpha)$.
Note that in the momentum representation $\alpha=p^{2}/p^{2}_{max}$,
$p_{max}\sim p_{pl}$,where $p_{pl}$ is the Planck momentum.
When present in matrix elements, $exp(-\alpha)$ can damp the
contribution of great momenta in a perturbation theory. The solution
(\ref{U1})given by the formula (\ref{U4}) is further referred to as
{\bf(exponential ansatz)}.This ansatz will be the principal one in  our
further consideration.
\end{enumerate}
In\cite{r9},\cite{r10} it has been demonstrated, how a transition
from Neumann's picture to Shr{\"o}dinger's picture, i.e. from the
density matrix deformation to the wave function deformation, may
be realized by the proposed approach
\\ \noindent {\bf Definition
2.} {\bf(Quantum Mechanics with Fundamental Length
[Shr{\"o}dinger's picture])}
\\
\\
Here, the prototype of Quantum Mechanical normed wave function (or
the pure state prototype) $\psi(q)$ with
$\int|\psi(q)|^{2}dq=1$ in QMFL is $\theta(\alpha)\psi(q)$. The
parameter of deformation $\alpha$ assumes the value
$0<\alpha\leq1/4$. Its properties are
$|\theta(\alpha)|^{2}<1$,$\lim\limits_{\alpha\rightarrow
0}|\theta(\alpha)|^{2}=1$ and the relation
$|\theta(\alpha)|^{2}-|\theta(\alpha)|^{4}\approx \alpha$ takes
place. In such a way the total probability always is less than 1:
$p(\alpha)=|\theta(\alpha)|^{2}=\int|\theta(\alpha)|^{2}|\psi(q)|^{2}dq<1$
tending to 1 when  $\alpha\rightarrow 0$. In the most general
case of the arbitrarily normed state in QMFL(mixed state prototype)
$\psi=\psi(\alpha,q)=\sum_{n}a_{n}\theta_{n}(\alpha)\psi_{n}(q)$
with $\sum_{n}|a_{n}|^{2}=1$ the total probability is
$p(\alpha)=\sum_{n}|a_{n}|^{2}|\theta_{n}(\alpha)|^{2}<1$ and
 $\lim\limits_{\alpha\rightarrow 0}p(\alpha)=1$.

It is natural that Shr{\"o}dinger equation is also deformed in
QMFL. It is replaced by the equation

\begin{equation}\label{U24}
\frac{\partial\psi(\alpha,q)}{\partial t}
=\frac{\partial[\theta(\alpha)\psi(q)]}{\partial
t}=\frac{\partial\theta(\alpha)}{\partial
t}\psi(q)+\theta(\alpha)\frac{\partial\psi(q)}{\partial t},
\end{equation}
where the second term in the right-hand side generates the
Shr{\"o}dinger equation as
\begin{equation}\label{U25}
\theta(\alpha)\frac{\partial\psi(q)}{\partial
t}=\frac{-i\theta(\alpha)}{\hbar}H\psi(q).
\end{equation}

Here $H$ is the Hamiltonian and the first member is added
similarly to the member that appears in the deformed Liouville
equation, vanishing when $\theta[\alpha(t)]\approx const$. In
particular, this takes place in the low energy limit in QM, when
$\alpha\rightarrow 0$. It should be noted that the above theory is
not a time reversal of QM because the combination
$\theta(\alpha)\psi(q)$ breaks down this property in the deformed
Shr{\"o}dinger equation. Time-reversal is conserved only in the
low energy limit, when a quantum mechanical Shr{\"o}dinger
equation is valid.
\section{Some Comments and Unitarity in QMFL}
As has been indicated in the previous section, time reversal is
retained in the large-scale limit only. The same is true for the
superposition principle in Quantum Mechanics. Indeed, it may be
retained in a very narrow interval of cases for the functions
$\psi_{1}(\alpha,q)=\theta(\alpha)\psi_{1}(q)$ è
$\psi_{2}(\alpha,q)=\theta(\alpha)\psi_{2}(q)$ with the same value
$\theta(\alpha)$. However, as for all $\theta(\alpha)$, their
limit is $\lim\limits_{\alpha\rightarrow 0}|\theta(\alpha)|^{2}=1$
or equivalently $\lim\limits_{\alpha\rightarrow
0}|\theta(\alpha)|=1$, in going to the low-energy limit each wave
function $\psi(q)$ is simply multiplied by the phase factor
$\theta(0)$. As a result we have Hilbert Space wave functions in
QM. Comparison of both pictures (Neumann's and Shr{\"o}dinger's)
is indicative of the fact that unitarity means the retention of
the probabilities $\omega_{i}(\alpha)$ or retention of the squared
modulus (and hence the modulus) for the function $\theta(\alpha)$:
$|\theta(\alpha)|^{2}$,($|\theta(\alpha)|$).That is
\\
\\
$$\frac{d\omega_{i}[\alpha(t)]}{dt}=0$$ or
$$\frac{d|\theta[\alpha(t)]|}{dt}=0.$$
\\
\\
In this way a set of unitary transformations
of QMFL includes a group
$U$ of the unitary transformations for the wave functions
$\psi(q)$ in QM.
\\It is seen that on going from Planck's scale to the
conventional one , i.e. on transition from the Early Universe to
the current one, the scale has been rapidly changing in the
process of inflation expansion and the above conditions
failed to be fulfilled:
\begin{equation}\label{U26}
\frac{d\omega_{i}[\alpha(t)]}{dt}\neq 0, {\sloppy}
\frac{d|\theta[\alpha(t)]|}{dt}\neq 0.
\end{equation}
In terms of the density pro-matrices of section 2 this is a
limiting transition from the density pro-matrix in QMFL
$\rho(\alpha)$,$\alpha>0$ , that is a prototype of the pure state
at $\alpha\rightarrow 0$, to the density matrix $\rho(0)=\rho$
representing a pure state in QM. Mathematically this means that a
nontotal probability (below 1) is changed by the total one (equal
to 1). For the wave functions in Schr{\"o}dinger picture this
limiting transition from QMFL to QM is as follows:
\\
\\
$$\lim\limits_{\alpha\rightarrow 0}\theta(\alpha)\psi(q)=\psi(q)$$
up to the phase factor.
\\It is apparent that the above transition from QMFL to QM is
not a unitary process, as indicated in
\cite{r3},\cite{r4},\cite{r8}-\cite{r10}.
However, the unitarity may be recovered when we consider
in a sense a reverse process:
absorption of the matter by a black hole and its transition to singularity
conforming to the reverse and nonunitary transition from QM to QMFL.
Thus, nonunitary transitions occur in this picture twice:
\\
\\
$$I.(QMFL,OS,\alpha\approx 1/4)\stackrel{Big\enskip
Bang}{\longrightarrow}(QM,\alpha\approx 0)$$
\\
\\
$$II.(QM,\alpha\approx 0)\stackrel{absorbing\enskip BH
}{\longrightarrow}(QMFL,SBH,\alpha\approx 1/4),$$
\\
\\
Here the following abbreviations are used:
OS for the Origin Singularity; BH for a Black
Hole; SBH for the Singularity in Black Hole.
\\
As a result of these two nonunitary transitions, the total
unitarity may be recovered:
\\
\\
$$III.(QMFL,OS,\alpha\approx
1/4){\longrightarrow}(QMFL,SBH,\alpha\approx 1/4)$$
\\
\\
In such a manner the total information quantity in the Universe
remains unchanged, i.e. no information loss occurs.
\\ In terms of the deformed Liouville equation \cite{r8}-\cite{r10}
we arrive to the expression with the same right-hand parts for
$t_{initial}\sim t_{Planck}$ and $t_{final}$  (for $\alpha\approx
1/4$).
\begin{eqnarray}\label{U27}
\frac{d\rho[\alpha(t),t]}{dt}=\sum_{i}
\frac{d\omega_{i}[\alpha(t)]}{dt}|i(t)><i(t)|-\nonumber \\
-i[H,\rho(\alpha)]= d[ln\omega(\alpha)]\rho
(\alpha)-i[H,\rho(\alpha)].
\end{eqnarray}
It should be noted that for the closed Universe one can consider
Final Singularity (FS) rather than the Singularity of Black Hole
(SBH), and then the right-hand parts of diagrams II and III
will be changed:
\\
\\
$$IIa.(QM,\alpha\approx 0)\stackrel{Big\enskip Crunch
}{\longrightarrow}(QMFL,FS,\alpha\approx 1/4),$$
\\
\\
$$IIIa.(QMFL,OS,\alpha\approx
1/4){\longrightarrow}(QMFL,FS,\alpha\approx 1/4)$$
\\
\\
At the same time, in this case the general unitarity and information
are still retained, i.e. we again have the "unitary" product of
two "nonunitary" arrows:
\\
\\
$$IV.(QMFL,OS,\alpha\approx 1/4)\stackrel{Big\enskip
Bang}{\longrightarrow}(QM,\alpha\approx 0)\stackrel{Big\enskip
Crunch }{\longrightarrow}(QMFL,FS,\alpha\approx 1/4),$$
\\
\\
Finally, arrow III may appear immediately, i.e. without the
appearance of arrows I è II, when in the Early Universe
mini BH are arising:
\\
\\
$$IIIb.(QMFL,OS,\alpha\approx
1/4){\longrightarrow}(QMFL, mini\enskip BH, SBH,\alpha\approx
1/4)$$
\\
\\
Note that here, unlike the previous cases, a unitary transition
occurs immediately, without any additional nonunitary ones, and
with retention of the total information.
\\ Another approach to the information paradox problem associated
with the above-mentioned methods (density matrix deformation) is
the introduction and investigation of a new value namely: entropy
density per minimum unit area. This approach is described in
section 4.

\section{ Entropy Density Matrix and Information Loss Problem }

In \cite{r3},\cite{r4},\cite{r8},\cite{r9},\cite{r10} the authors
were too careful, when introducing for density pro-matrix
$\rho(\alpha)$ the value $S_{\alpha}$ generalizing the ordinary
statistical entropy:
\\
 $$S_{\alpha}=-Sp[\rho(\alpha)\ln(\rho(\alpha))]=
 -<\ln(\rho(\alpha))>_{\alpha}.$$
\\
In \cite{r9},\cite{r10} it was noted that $S_{\alpha}$ means
the entropy density   on a minimum unit area depending on the
scale. In fact a more general concept accepts the form of the
entropy density matrix \cite{r11}:
\begin{equation}\label{U4a}
S^{\alpha_{1}}_{\alpha_{2}}=-Sp[\rho(\alpha_{1})\ln(\rho(\alpha_{2}))]=
-<\ln(\rho(\alpha_{2}))>_{\alpha_{1}},
\end{equation}
where $0< \alpha_{1},\alpha_{2}\leq 1/4.$
\\ $S^{\alpha_{1}}_{\alpha_{2}}$ has a clear physical meaning:
the entropy density is computed  on the scale associated with the
deformation parameter $\alpha_{2}$ by the observer who is at a
scale corresponding to the deformation parameter $\alpha_{1}$.
Note that with this approach the diagonal element
$S_{\alpha}=S_{\alpha}^{\alpha}$,of the described matrix
$S^{\alpha_{1}}_{\alpha_{2}}$ is the density of entropy measured
by the observer  who is at the same scale  as the measured object
associated with the deformation parameter $\alpha$. In \cite{r10}
section 6 such a construction was used implicitly in derivation of
the semiclassical Bekenstein-Hawking formula for the Black Hole
entropy:

a) For the initial (approximately pure) state
\\
$$S_{in}=S_{0}^{0}=0$$
\\
b) Using the exponential ansatz(\ref{U4}),we obtain:
\\
$$S_{out}=S^{0}_{\frac{1}{4}}=-<ln[exp(-1/4)]\rho_{pure}>=-<\ln(\rho(1/4))>
=\frac{1}{4}.$$
\\
So increase in the entropy density for an external observer at the
large-scale limit is 1/4. Note that increase of the entropy
density(information loss) for the observer crossing the
horizon of the black hole's events and moving with the information
flow to singularity will be smaller:
\\
$$S_{out}=S_{\frac{1}{4}}^{\frac{1}{4}}=-Sp(exp(-1/4)
ln[exp(-1/4)]\rho_{pure})=-<\ln(\rho(1/4))>_{\frac{1}{4}} \approx
0.1947 .$$ It is clear that this fact may be interpreted as
follows: for the observer moving together with information its
loss can  occur only at the transition to smaller scales, i.e. to
greater deformation parameter $\alpha$. \\
\\ Now we consider the general Information Problem.
Note that with the classical Quantum Mechanics (QM) the entropy
density matrix $S^{\alpha_{1}}_{\alpha_{2}}$ (\ref{U4a}) is
reduced only to one element $S_{0}^{0}$ . So we can not test
anything. Moreover, in previous works relating the quantum
mechanics of black holes and information paradox
\cite{r16},\cite{r17,r18} the initial and final states when a
particle hits the
hole are treated proceeding from different theories
(QM and QMFL respectively), as was indicated in diagram II:
\\
\\
(Large-scale limit, QM,
 density matrix) $\rightarrow$ (Black Hole, singularity, QMFL,
density pro-matrix),
\\
\\
Of course in this case any conservation of information is
impossible as these theories are based on different concepts of
entropy. Simply saying, it is incorrect to compare the entropy
interpretations of two different theories (QM and QMFL)where this
notion is originally differently understood. So the chain above
must be symmetrized by accompaniment of the arrow on the left ,so
in an ordinary situation we have a chain (diagram III):
\\
\\
(Early Universe, origin singularity, QMFL, density pro-matrix)
$\rightarrow$
\\ (Large-scale limit, QM,
 density matrix)$\rightarrow$ (Black Hole, singularity, QMFL,
density pro-matrix),
\\
\\
So it's more correct to compare entropy close to the origin and
final (Black hole) singularities. In other words, it is necessary
to take into account not only the state, where information
disappears, but also that whence it appears. The question arises,
whether in this case the information is lost for every separate
observer. For the event under consideration this question sounds
as follows: are the entropy densities S(in) and S(out) equal for
every separate observer? It will be shown that in all conceivable
cases they are equal.

1) For the observer in the large-scale limit (producing
measurements in the semiclassical approximation) $\alpha_{1}=0$
\\
\\
$S(in)=S^{0}_{\frac{1}{4}}$ (Origin singularity)
\\
\\
$S(out)=S^{0}_{\frac{1}{4}}$ (Singularity in Black Hole)
\\
\\
So $S(in)=S(out)=S^{0}_{\frac{1}{4}}$. Consequently, the initial
and final densities of entropy are equal and there is no
information loss.
\\
2) For the observer moving together with the information flow in
the general situation  we have the chain:
\\
$$S(in)\rightarrow S(large-scale)\rightarrow S(out),$$
\\
where $S(large-scale)=S^{0}_{0}=S$. Here $S$ is an ordinary
entropy of Quantum Mechanics(QM), but
$S(in)=S(out)=S^{\frac{1}{4}}_{\frac{1}{4}}$,- value considered in
QMFL. So in this case the initial and final densities of entropy
are equal without any loss of information.
\\
3) This case is a special case of 2), when we do not come out of
the Early Universe considering the processes with the
participation of black mini-holes only. In this case the
originally specified chain becomes shorter by one section (diagram
IIIb):
\\
\\
(Early Universe, origin singularity, QMFL, density
pro-matrix)$\rightarrow$ (Black Mini-Hole, singularity, QMFL,
density pro-matrix),
\\
\\
and member $S(large-scale)=S^{0}_{0}=S$ disappears at the
corresponding chain of the entropy density associated with the
large-scale consideration:
\\
$$S(in)\rightarrow S(out),$$
\\
It is, however, obvious that in case
$S(in)=S(out)=S^{\frac{1}{4}}_{\frac{1}{4}}$ the density of
entropy is preserved. Actually this event was mentioned in section
5 \cite{r10},where from the basic principles it has been found
that black mini-holes do not radiate, just in agreement with the
results of other authors \cite{r13}-\cite{r15},\cite{r19}
\\ As a result, it's possible to write briefly
\\
$$S(in)=S(out)=S^{\alpha}_{\frac{1}{4}},$$
\\
where $\alpha$ - any value in the interval $0<\alpha\leq 1/4.$
\\ Actually our inferences are similar to those of previous section
in terms of the Liouville's equation deformation
(\ref{U27}):
\\
$$\frac{d\rho}{dt}=\sum_{i}
\frac{d\omega_{i}[\alpha(t)]}{dt}|i(t)><i(t)|-i[H,\rho(\alpha)]=
\\d[ln\omega(\alpha)]\rho (\alpha)-i[H,\rho(\alpha)].$$
\\
The main result of this section is a necessity to account for the
member $d[ln\omega(\alpha)]\rho (\alpha)$,deforming the right-side
expression of $\alpha\approx 1/4$.

\section {Unitarity, Non-Unitarity and Heisenberg's Algebra Deformation}

The above-mentioned unitary and nonunitary transitions may be
described in terms of Heisenberg's algebra deformation
(deformation of commutators) as well. We use the principal results
and designations from \cite{r5}.In the process the following
assumptions are resultant: 1)The three-dimensional rotation group
is not deformed; angular momentum ${\bf J}$ satisfies the
undeformed $SU(2)$ commutation relations, whereas the coordinate and
momenta satisfy the undeformed commutation relations $\left[
J_i,x_j\right] =i\epsilon_{ijk}x_k, \left[ J_i,p_j\right]
=i\epsilon_{ijk}p_k$. 2) The momenta commute between themselves:
$\left[ p_i,p_j\right] =0$, so the translation group is also not
deformed. 3) Commutators $\left[ x,x\right]$ and $\left[ x,p\right]$
depend on the deformation parameter $\kappa$ with the
dimension of mass. In the limit $\kappa\rightarrow \infty$ with
$\kappa$ much larger than any energy the canonical commutation
relations are recovered.
\\
For a specific realization of points 1) to 3) the generating GUR
are of the form \cite{r5}: ($\kappa$-deformed Heisenberg algebra)
\begin{eqnarray}
\left[ x_i ,x_j \right] &= & -\frac{\hbar^2}{\kappa^2}\,
i\epsilon_{ijk}J_k\label{xx}\\ \left[ x_i , p_j \right]   &= &
i\hbar\delta_{ij} (1+\frac{E^2}{\kappa^2})^{1/2}\, .\label{xp}
\end{eqnarray}
Here $E^2=p^2+m^2$. Note that in this formalism
the transition from GUR to UR, or equally from
QMFL to QM with $\kappa\rightarrow
\infty$ or from Planck scale to the conventional one, is nonunitary
exactly following the transition from density pro-matrix to
the density matrix in previous sections:
\\
$$\rho(\alpha\neq 0)\stackrel{\alpha\rightarrow
0}{\longrightarrow}\rho$$
\\
Then the first arrow I in the formalism of this section
may be as follows:
\\
$$I^{\prime}.(GUR,OS,\kappa\sim M_{p})\stackrel{Big\enskip
Bang}{\longrightarrow}(UR,\kappa=\infty)$$ or what is the same
$$I^{\prime\prime}.(QMFL,OS,\kappa\sim M_{p})\stackrel{Big\enskip
Bang}{\longrightarrow}(QM,\kappa=\infty),$$
\\
where $M_{p}$ is the Planck mass. In some works of the last two
years Quantum Mechanics for a Black Hole has been already
considered as a Quantum Mechanics with GUR \cite{r13}-\cite{r15}.
As a consequence, by this approach the Black Hole is not
completely evaporated but rather some stable remnants always
remain in the process of its evaporation with a mass $\sim M_{p}$.
In terms of \cite{r5} this means nothing else but a reverse
transition: $(\kappa=\infty)\rightarrow(\kappa\sim M_{p})$. And
for an outside observer this transition is of the form:
\\
$$II^{\prime}.(UR,\kappa=\infty)\stackrel{absorbing\enskip
BH}{\longrightarrow}(GUR,SBH,\kappa\sim M_{p}),$$ òî åñòü
$$II^{\prime\prime}.(QM,\kappa=\infty)\stackrel{absorbing\enskip
BH}{\longrightarrow}(QMFL,SBH,\kappa\sim M_{p}).$$
\\
\\
So similar to the previous section, two nonunitary inverse transitions
a)$I^{\prime},(I^{\prime\prime})$ and b)$II^{\prime},(II^{\prime\prime})$
are liable to generate a unitary transition:
\\
$$III^{\prime}.(GUR,OS,\kappa\sim M_{p})\stackrel{Big\enskip
Bang}{\longrightarrow}(UR,\kappa=\infty)\stackrel{absorbing\enskip
BH}{\longrightarrow}(GUR,SBH,\kappa\sim M_{p}),$$
\\
or to summerize
\\
$$III^{\prime\prime}.(GUR,OS,\kappa\sim
M_{p})\rightarrow(GUR,SBH,\kappa\sim M_{p})$$
\\
In conclusion of this section it should be noted that
an approach to the Quantum Mechanics at Planck Scale
using the Heisenberg algebra deformation (similar to the approach
based on the density matrix deformation from the
previous section) gives a deeper insight into the possibility of
retaining the unitarity and the total quantity of information in
the Universe, making possible the solution of Hawking's information
paradox problem
\cite{r16}-\cite{r18}.

\section {Statistical Mechanics Deformation and Transitions}
Naturally, deformation of Quantum Mechanics in the Early Universe
is associated with the Statistical Mechanics deformation as indicated
in \cite{r12},\cite{r20}. In case under consideration this simply
implies a transition from the Generalized Uncertainty Relations (GUR)
of Quantum Mechanics to GUR in Thermodynamics \cite{r20}- \cite{r22}.
The latter are distinguished from the normal uncertainty relations by:
\begin{equation}\label{U1T}
\Delta \frac{1}{T}\geq\frac{k}{\Delta U}
\end{equation}
i.e. by inclusion of the high-temperature term into the right-hand side
\begin{equation}\label{U2T}
\Delta \frac{1}{T}\geq
  \frac{k}{\Delta U}+\alpha^{\prime}
  \frac{1}{T_{p}^2}\frac{\Delta U}{k}+...
\end{equation}
dots meaning the existence of higher order corrections
\cite{r20}. Thus, denoting the
Generalized Uncertainty Relations in Thermodynamics as GURT and
using abbreviation URT for the conventional ones, we obtain a new form
of diagram I from section III
($I^{\prime}$ of section IV respectively):
\\
 $$I^{T}.(GURT,OS)\stackrel{Big\enskip
Bang}{\longrightarrow}(URT)$$
\\
In \cite{r12},\cite{r20} the Statistical Mechanics deformation
associated with GURT is described by the introduction of
the respective deformation for the statistical density matrix
$\rho_{stat}(\tau)$ where $0<\tau \leq 1/4$.
Obviously, close to the Origin Singularity $\tau\approx
1/4$. Because of this, arrow $I^{T}$ may be represented
in a more general form as
\\
 $$I^{Stat}.(GURT,OS,\rho_{stat}(\tau),\tau\approx 1/4)
 \stackrel{Big\enskip
Bang}{\longrightarrow}(URT,\rho_{stat},\tau\approx 0)$$
\\
The reverse transition is also possible. In \cite{r13}-\cite{r15}
it has bee shown that Statistical Mechanics of Black Hole should
be consistent with the deformation of a well-known Statistical
Mechanics. The demonstration of an *upper* bound for temperature
in Nature, given by Planck temperature and related to Black Hole
evaporation, was provided in \cite{r23}. It is clear that
Emergence of such a high temperatures is due to GURT. And we have
the following diagram that is an analog of diagrams II and
$II^{\prime}$ for Statistical Mechanics:
\\
$$II^{Stat}.(URT,\rho_{stat},\tau\approx 0)
\stackrel{absorbing\enskip
BH}{\longrightarrow}(GURT,SBH,\rho_{stat}(\tau),\tau\approx 1/4).
$$
\\
By this means, combining $I^{Stat}$ and $II^{Stat}$, we obtain
$III^{Stat}$ representing a complete statistical-mechanics analog
for quantum-mechanics diagrams $III$ and $III^{\prime}$:
\\
$$III^{Stat}.(GURT,OS,\tau\approx 1/4)
 \stackrel{Big\enskip
Bang,\enskip absorbing\enskip BH}{\longrightarrow}
(GURT, SBH,\tau\approx 1/4).$$
\\
And in this case two nonunitary transitions $I^{Stat}$
and $II^{Stat}$ in the end lead to a unitary transition
$III^{Stat}$.

\section{Measuring Procedure and New Small Parameter}
As noted above, the primary relation may be written
in the form of a series \begin{equation}\label{U1b}
Sp[\rho(\alpha)]-Sp^{2}[\rho(\alpha)]=\alpha+a_{0}\alpha^{2}
+a_{1}\alpha^{3}+...
\end{equation}
As a result, a measurement procedure using the exponential ansatz
(\ref{U4}) may be understood as the calculation of factors
$a_{0}$,$a_{1}$,... or the definition of additional members in the
exponent "destroying" $a_{0}$,$a_{1}$,... . It is easy to check
that the exponential ansatz gives $a_{0}=-3/2$, being coincident
with the logarithmic correction factor for the Black Hole entropy
\cite{r24}.
\\From section 2 and specifically from relation (\ref{U1b})
it follows that $\alpha$ is a new small parameter.
Among its obvious advantages one could name:
\\1)  its dimensionless nature,
\\2)  its variability over the finite interval $0<\alpha \leq 1/4$.
Besides, for the well-known physics
it is actually very small:
$\alpha\sim 10^{-66+2n}$, where $10^{-n}$ is the measuring
scale. Here the Planck scale $\sim 10^{-33}cm$ is assumed;
\\3)and finally the calculation of this parameter involves all
three fundamental constants, since by Definition 1 of section 2
$\alpha = l_{min}^{2}/x^{2 }$, where $x$ is the measuring scale
and $l_{min}^{2}\sim l_{pl}^{2}=G\hbar/c^{3}$.
\\ Therefore, series expansion in $\alpha$ may be of great importance.
Especially as all field constants of any quantum system by
Definition 2 of section 2 are dependent on $\alpha$, i.e.
$\psi=\psi(\alpha)$.

\section{Conclusion}

Thus, this work outlines that the existence of GUR and hence the
appearance of QMFL enables a better understanding of the
information problem in the Universe providing a key to the
solution of this problem in a not inconsistent manner, practically
in the same way but irrespective of the approach used: 1) density
matrix deformation in Quantum (Statistical) Mechanics at Planck's
scale (and as a consequence,  entropy density matrix approach)
or 2) Heisenberg algebra deformation.
\\
It should be noted that the question of the relationship between these two
approaches, i.e. transition from one deformation to the other,
still remains open. This aspect is to be studied in further
investigations of the author.

%References

\end{document}